\documentstyle[12pt]{article}

\def\pre#1#2#3{Phys. Rev. E, {\bf #1}, #2 (#3)}

\def\phd#1#2#3{Physica D {\bf #1}, #2 (#3)}

\def\beq{\begin{equation}}
\def\eqn{\end{equation}\noindent}
\def\LEE{\lambda}
\def\LESTAR{$\LEE^{\star}$}
\def\LE{$\LEE$~}
\def\LEc{$\LEE$,~}

\begin{document}
\begin{center}
{\Large{\bf Targetting Chaos through Adaptive Control}}\\ 
\vspace{1cm}
Ramakrishna Ramaswamy $^{a,}$\footnote{{rama@jnuniv.ernet.in}, $^2$
{sudeshna@imsc.ernet.in}, $^3$ {gupte@chaos.iitm.ernet.in}},
Sudeshna Sinha$^{b,2}$ and Neelima Gupte$^{c,3}$\\ 
{\it $^a$ School of Physical Sciences, Jawaharlal Nehru University,\\ New
Delhi 110067, India}\\ $^b$ {\it Institute of Mathematical Sciences,
Taramani, Chennai 600 113, India}\\ $^c$ {\it Department of Physics,
IIT Madras, Chennai 600036, India}
\date{\today}
\end{center}
\begin{abstract} We describe adaptive control algorithms whereby a
chaotic dynamical system can be steered to a target state with 
desired characteristics. A specific implementation considered has the
objective of directing the system to a state which is more chaotic or
mixed than the uncontrolled one. This methodology is easy to implement
in discrete or continuous dynamical systems. It is robust and
efficient, and has the additional advantage that knowledge of the
detailed behaviour of the system is not required.
\end{abstract}
\vskip1cm

Adaptive control algorithms have hitherto been implemented for the
purpose of maintaining periodic behaviour in nonlinear systems
\cite{HL,SRS,GENERAL}. Recently there has been some interest in control
algorithms whose aim is to target other types of dynamical behaviour.
``Anti--control'' algorithms, namely those wherein the objective is to
maintain \cite{MAINT} or to {\em enhance} \cite{ENHANCE} the chaoticity
of dynamical systems have been devised. These efforts have been
motivated by practical situations where the enhancement or maintenance
of chaos has desirable consequences.  Examples of these can be found in
contexts as diverse as mixing flows \cite{O}, electronic systems
\cite{OD} and chemical reactions \cite{CR}, where the enhancement of
chaos can lead to improved performance, or in biological applications
such as in neural systems, where the maintenance of chaos provides the
key to the avoidance of pathological behaviour \cite{SJDCSD}.

In this paper we describe an adaptive anti--control algorithm which is
simple and easily implemented. The algorithm is set up to maintain a
desired level of chaoticity, and to achieve a target value of the local
Lyapunov exponent or a local stretching rate. The technique is
sufficiently general and can be extended so as to make a given
dynamical system achieve a target value of any desired variable or
function.

In the context of nonlinear dynamical systems, the method of adaptive
control \cite{HL,SRS} applies a feedback loop in order to drive the
system parameter (or parameters) to the values required so as to
achieve a desired or target state. This is implemented by augmenting
the evolution equation for the dynamical system by an additional
equation for the evolution of the parameter(s) as described below.

Consider a general $N$-dimensional dynamical system described by the
evolution equation
\beq
\dot{\bf X} = {\bf F} ({\bf X}; \mu; t)
\eqn
where ${\bf X} \equiv (X_1, X_2, \dots X_N)$ are the state variables
and $\mu$ is the parameter whose value determines the nature of the
dynamics.  The adaptive control is effected by the additional dynamics
\beq 
\label{ds}
\dot \mu = \epsilon ( {\cal P}^{\star} - {\cal P}) 
\eqn 
where ${\cal P}^{\star}$ is the target value of some variable or
property (which could be a function of several variables) ${\cal P}$,
and the value of $\epsilon$ indicates the stiffness of control. The
extension to the situation of several control parameters is
straightforward.

The scheme is adaptive since in the above procedure the parameters
which determine the nature of the dynamics {\it self-adjust} or adapt
themselves to yield the desired dynamics. This has also been termed
``dynamic feedback control'' in the literature \cite{S}. The adaptive
principle is remarkably robust and efficient in generic nonlinear
systems \cite{SRS}, and may therefore be of considerable utility in a
large variety of phenomena, ranging from biological units to control
engineering.

For the maintenance of a stable fixed point \cite{HL} in a discrete
dynamical system for example, the procedure is as follows.  The
nonlinear system evolves according to the appropriate equation,
\beq
x_{n+1} = f(\alpha,x_n)
\label{dls}
\eqn
where $\alpha$ is the parameter to be controlled. If the required value
of $x$ is, say, $x^{\star}$, then the additional equation (for ${\cal
P} \equiv x$)
\beq
\alpha_{n+1} = \alpha_n + \epsilon (x^{\star} - x_n)
\label{control}
\eqn
has the desired effect of tuning the value of $\alpha$ so that the
dynamics of the combined equations gives $x \to x^{\star}$ over a wide
range of initial conditions. The stiffness $\epsilon$ determines how
rapidly the system is controlled. The control time, defined as the time
required to reach the desired state, is crucially dependent on the
value of $\epsilon$. Numerical experiments show that for small
$\epsilon$ the recovery time is inversely proportional to the stiffness
of control.  This follows from the fact that when $\epsilon$ is small
compared to the inverse timescales in the original dynamical system, we
can use an adiabatic approximation since $\dot{\mu} \to 0$, from which
\cite{S} it follows that control time will be proportional to
$1/{\epsilon}$. 

With modifications, this method can be made to control to a stable or
unstable periodic orbit of arbitrary period \cite{SRS,S,AJ}.

If the desired target state is chaotic rather than periodic, one needs
to choose an appropriate property $ {\cal P}$ which should reflect the
desired chaotic nature of the target state. Therefore the natural
choice of $ {\cal P} $ is the Lyapunov exponent.  It is thus clear
that in order to achieve a desired value of the Lyapunov exponent, say
\LESTAR, the procedure to be followed is similar (with ${\cal P}
\equiv \lambda$).

For a 1--$d$ discrete dynamical system as in 
Eq.~(\ref{dls}) above, the Lyapunov exponent is defined through,
\beq
\LEE = \lim_{N \to \infty} {1 \over N}\sum_{i=0}^{N-1} \ln 
\vert f^{\prime}(\alpha, x_i) \vert .
\eqn
The control equation (\ref{control}) takes the form
\beq
\alpha_{n+1} = \alpha_n + \epsilon ( \LEE^{\star} - \lambda_n ),
\eqn
where $\lambda_n = \ln\vert f^{\prime}(\alpha, x_n)\vert$ is the
instantaneous value of the Lyapunov exponent. Implementation of the
methodology in, say, the logistic equation, is direct and the relevant
equations are
\begin{eqnarray}
x_{n+1} &=& \alpha_n x_n ( 1 - x_n)\\
\lambda_n &=& \ln\vert \alpha_n (1 - 2 x_n) \vert\label{osl}\\
\alpha_{n+1} &=& \alpha_n + \epsilon (\LEE^{\star} - \lambda_n).
\end{eqnarray}

Shown in Fig.~1 is an implementation of the control for $\LEE^{\star}$
= 0.36. Since $\lambda(\alpha)$ for the logistic equation is a highly
nonmonotonic function, there can be several parameter values for which
the system has the same \LEc namely several different attractors with
the same Lyapunov exponent are possible.  For example, the Lyapunov
exponent is approximately 0.36 at $\alpha \approx 3.7$ as well as at
$\alpha \approx 3.86$.  Which of these values the system adaptively
goes to depends on the initial state, the stiffness and the effective
basin of attraction (in parameter space).  For small stiffness, the
system sticks closely to one or the other attractor, only occasionally
making an excursion from one to the other, while for large stiffness,
the fluctuations in the parameter are much larger, as can be seen in
Fig.~1. For small values of stiffness the time taken to reach the
desired goal is usually inversely proportional to the stiffness of
control. Note however that increasing stiffness beyond a point can make
the method unstable and the dynamics unbounded.  There is, therefore,
an optimal strategy to be employed. While the optimal strategy to be
used can be worked out easily in a practical implementation, an
analytic optimality criterion is difficult to define.

The distribution of finite-time Lyapunov exponents shows that the
short-time chaoticity properties of the adaptive system can be quite
different from the equivalent chaotic system. Shown in Fig.~2 
are the distributions of adaptively controlled systems with the above
average \LE = 0.36, and with different stiffness, for 20 and 50 steps
respectively. While the desired \LE is maintained in all cases, it is
clear that the adaptation works differently for large or small stiffness.
Low stiffness allows the system to explore different
attractors with different properties, giving a wider spread in the
Lyapunov exponents, while a higher stiffness ensures that the
local \LE $\sim$ \LESTAR, narrowing the distribution.

Extensions of this procedure can be made to control higher dimensional
systems.  

The fact that the control is always operative means that the augmented
system is robust to perturbations. Indeed, if the parameter is
perturbed to a very different value, the system readily and rapidly
recovers to a dynamics such that the Lyapunov exponent is (nearly)
\LESTAR, again with time that inversely depends on $\epsilon$.
Note however that this control works only in the case of positive
\LESTAR: one cannot adaptively control in this manner to a
periodic orbit.

An application of practical importance is in enhancing the mixing in
chaotic systems. The appropriate adaptive strategy then is to take ${\cal
P}$ to be the stretching rate. Eq.~(\ref{ds}) thus becomes
\beq
\dot \mu = \epsilon ( {\cal R}_{target} - {\cal R}_{local}),
\label{mudot}
\eqn
where ${\cal R}_{local}$ is the instantaneous local stretching rate, and
${\cal R}_{target}$ is the prescribed desired stretching rate, which can
in principle be in any one of the dynamical variables characterizing the
system.

As an example, consider the Lorenz attractor,
\begin{eqnarray}
\dot {x} &=& \sigma ( y - x ) \nonumber\\
\dot {y} &=& \mu x - y - x z\nonumber\\
\dot {z} &=& x y - b z
\label{Lor}
\end{eqnarray}
Choosing the evolution equation for the parameter to be
\beq
\dot \mu=\epsilon ( {\cal R}_{target} - \dot x)
\label{mucont}
\eqn
where the instantaneous stretching rate ${\cal R}_{local} = \dot x$,
with $\dot x$ given by the evolution equation (\ref{Lor}), achieves the
objective. Note that instead of the $x$ direction, $y$ or $z$ can
equally effectively be used in the above control.

In the absence of knowledge of the evolution equation the above
control can be effected by the discrete evolution equation
\begin{equation}
\mu_{t+\delta t}= \mu_{t} + \epsilon ({\cal R}^{\prime}_{target} -\Delta x_t)
\label{mudiscrete}
\eqn
where $\Delta x_t $ is the local stretching given by $\| x_t
-x_{t-\delta t}\|$ ($\delta t$ small) where ${\cal
R}^{\prime}_{target}= \delta t \times {\cal R}_{target}$.

Shown in Fig.~3 is the result of an implementation of this adaptive
anti--control equation (\ref{mudiscrete}) where the target stretching
rate is specified to be $1.0$ with the control stiffness
$\epsilon=0.1$.  As can be seen, the controlled parameter first rapidly
climbs (the rate of ascent being directly proportional to $1/\epsilon$)
from an initial value $\mu =35.0$. Around $\mu \sim 380$ , it settles
into fluctuations which are of the integrated
white noise type ({\it i.e.} the power spectrum of the time series of
these fluctuations is clearly $S(f) \sim 1/f^2$) lead to a very mixed
system. Starting off with any other value of $\mu$ leads to the same
result, as does control via equation (\ref{mucont}) using the relation
between ${\cal R}_{target}$ and ${\cal R}^{\prime}_{target}$ defined
above.

The nature of adaptive anti--control is such that the dynamics that
obtains is intrinsically mixing. In contrast with similar techniques
where a stable state is targetted, the present mechanism \cite{SIGN}
essentially drags the system rapidly to the first appropriate state
encountered in parameter space, namely one which matches the targetted
local stretching rate. The system, in effect moves from (chaotic)
attractor to attractor, with significant fluctuations in the parameter
that is being controlled. By choosing a small target stretching rate,
this same control mechanism can be used to obtain cycles as well.
Somewhat fuzzy cycles are obtained, with the fuzziness decreasing with
decreasing $\epsilon$.

It should be emphasized that knowledge of the map in the control
algorithm above is in principle not required, since the necessary
information required to implement adaptive anti--control is simply the
difference between the current value of the variable and its previous
value. On the other hand, it is essential that the parameter $\mu$
being controlled must have the driving power in order to effect large
dynamical changes.  Parameters which are suitable for controlling are
easy to identify through the appropriate dynamical phase diagrams.

In summary, we have presented here an adaptive algorithm which can be
used to achieve desired chaotic behaviour in nonlinear dynamical
systems. The anti--control technique, which is rapid, powerful and
robust, extends adaptive control methods for obtaining periodic orbits
\cite{HL,SRS,AJ}. We have applied this to the case of achieving a
target value of the Lyapunov exponent, or a desired value of the local
streching rate and found that the methodology is successful in a number
of examples, including multidimensional and multiparameter systems.

An important consideration is that the present method can be
implemented without explicit knowledge of the dynamics. The possibility
of treating the system as a black--box is likely to be of utility in
complex experimental applications \cite{CR,SJDCSD} necessitating the
controlled maintainence or enhancement of chaos.

\vskip1cm
{\sc Acknowledgments:} RR and NG would like to thank the ICTP, Trieste,
where some of this work was done, for hospitality and also acknowledge
the support of the Department of Science and Technology (Grant
SP/MO-5/92 and Grant SP/S2/E-03/96) respectively.

\newpage
\begin{center}
{\bf Figure Captions}
\end{center}
\begin{itemize}
\item[Fig. 1]{
Variation of the parameter $\alpha$ as a function of
iteration step. The target Lyapunov exponent is $\LEE^{\star} = 0.36$,
and the stiffness is a) $\epsilon = $ 0.001, and b) $\epsilon = $ 0.01.
The different curves correspond to different initial $\alpha$.  In all
cases, the target $\LEE^{\star}$ is achieved rapidly and maintained.}
\label{Fig 1.}

\item[Fig. 2]{
Probability distribution for finite time or $N$--step Lyapunov
exponents in the adaptively controlled system. The target Lyapunov
exponent is $\LEE^{\star}$ = 0.36 as in Fig.~1. a) $N$ = 20 and b) $N$
= 50 correspond to a stiffness of $\epsilon = $ 0.001, while c) $N$ =
20 and d) $N$ = 50 correspond to a stiffness of $\epsilon = $ 0.01. }
\label{Fig 2.}

\item[Fig. 3] {
Variation of the parameter $\mu$ as a function of time for
the Lorenz attractor. The target stretching rate is $1.0$, the
stiffness is $\epsilon=0.1$ and the time step is $dt=0.01$.}
\label{Fig 3.}
\end{itemize}
 

\begin{thebibliography}{99}
\bibitem{SIGN} While the sign of the control dynamics given above holds
in generic cases, there may be situations where the sign of the control
dynamics is negative (for instance in the case of parameters $\sigma$
and $b$ in the Lorenz system \cite{OTT}). This will happen if the value
of the initial parameter is larger than the closest parameter which
matches the target (not smaller, as is usual). In real implementations
(where the dynamics is a black box) a gross scan of the parameter space
will give a reasonable indication as to which sign to use.
\end{thebibliography}
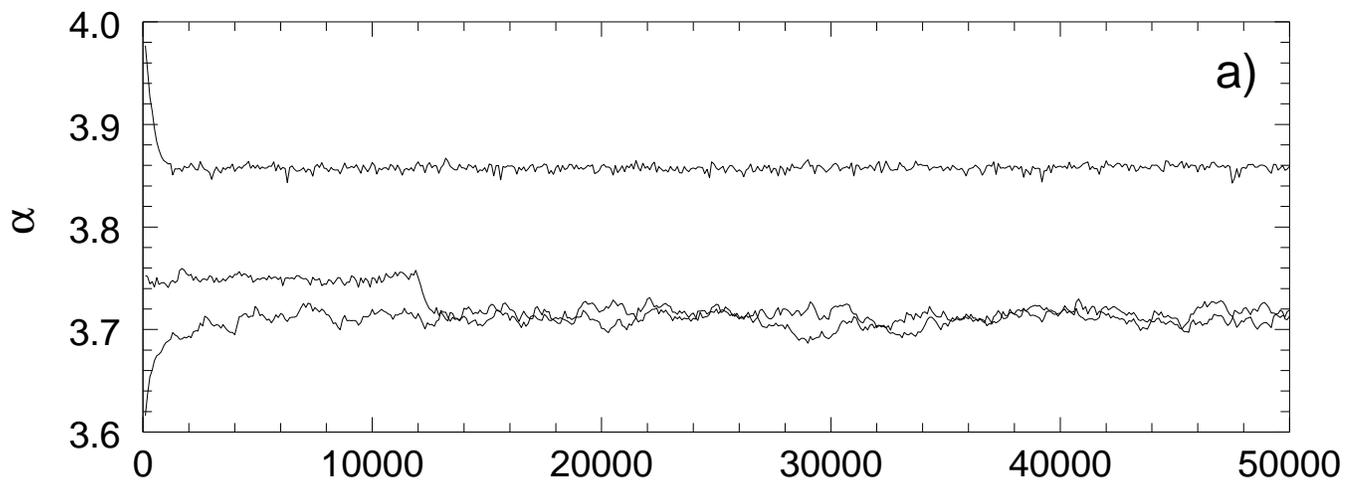
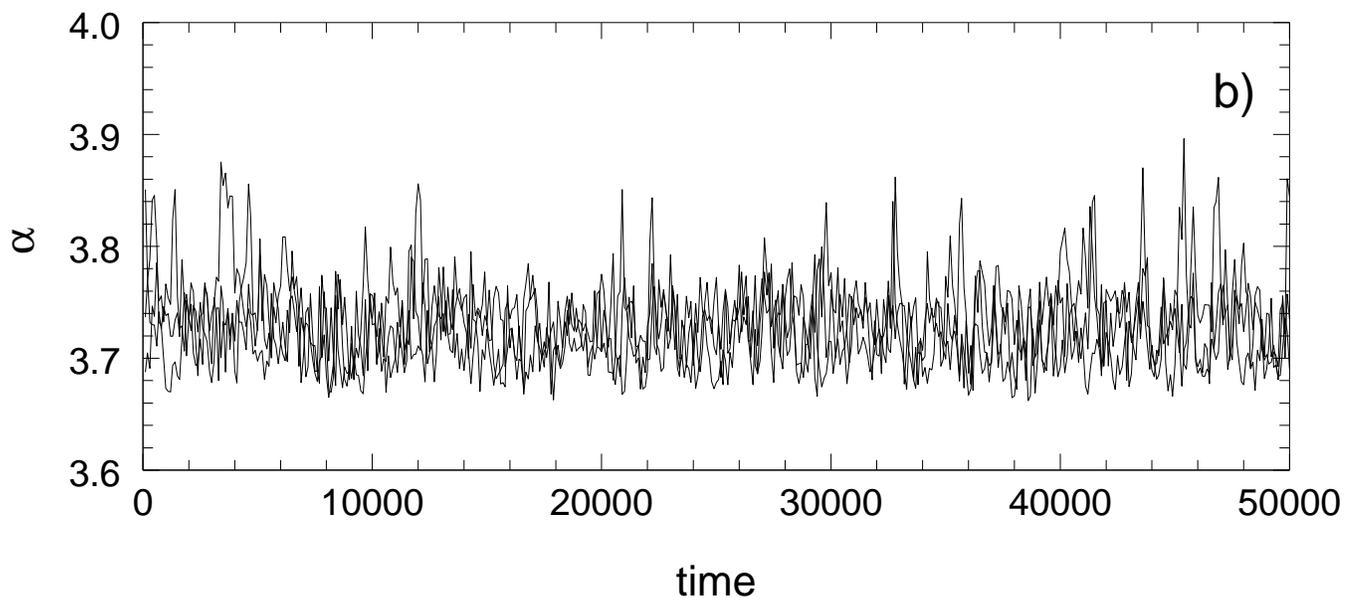
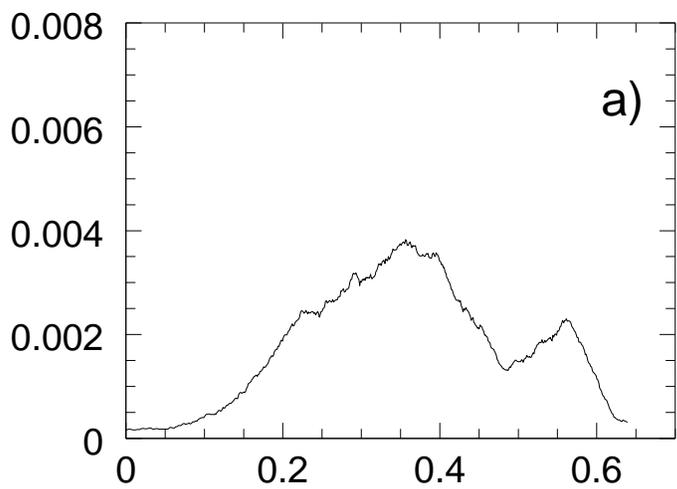
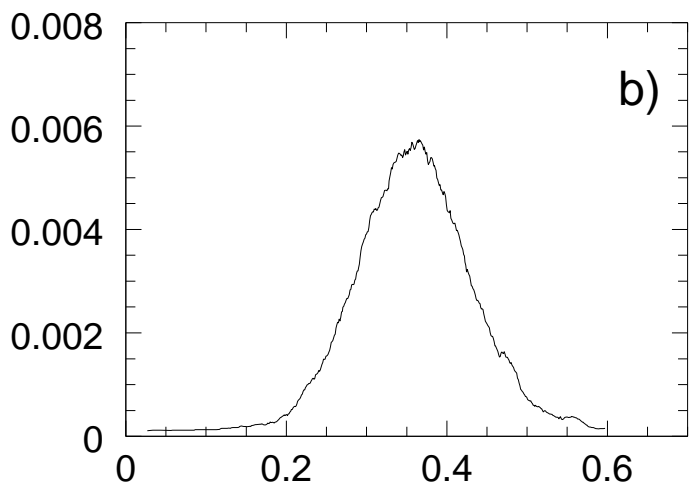
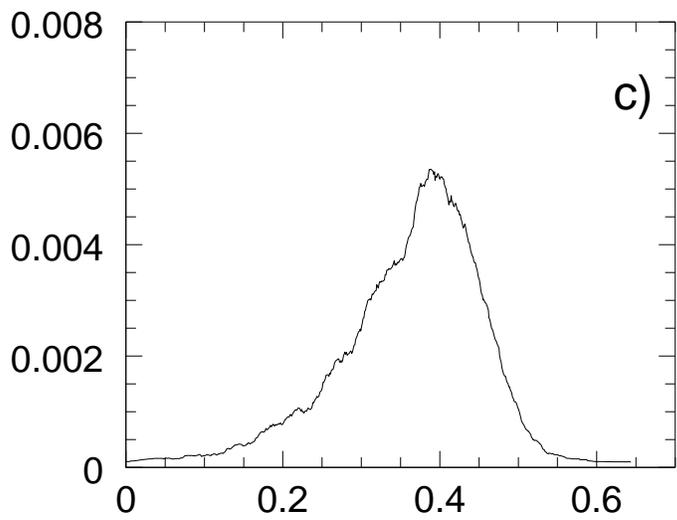
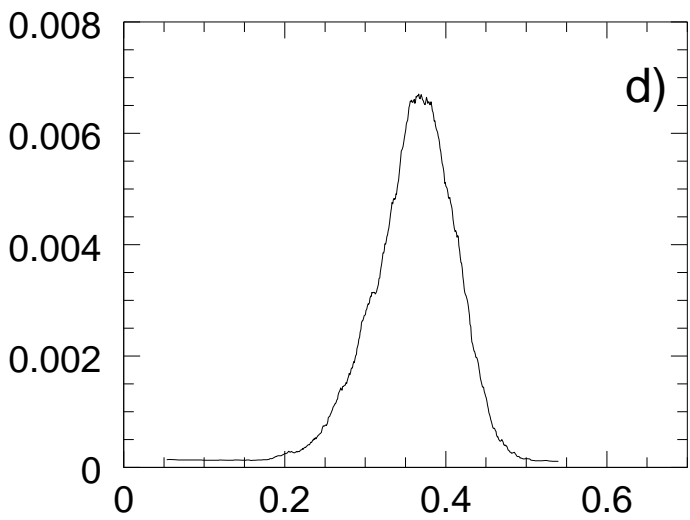
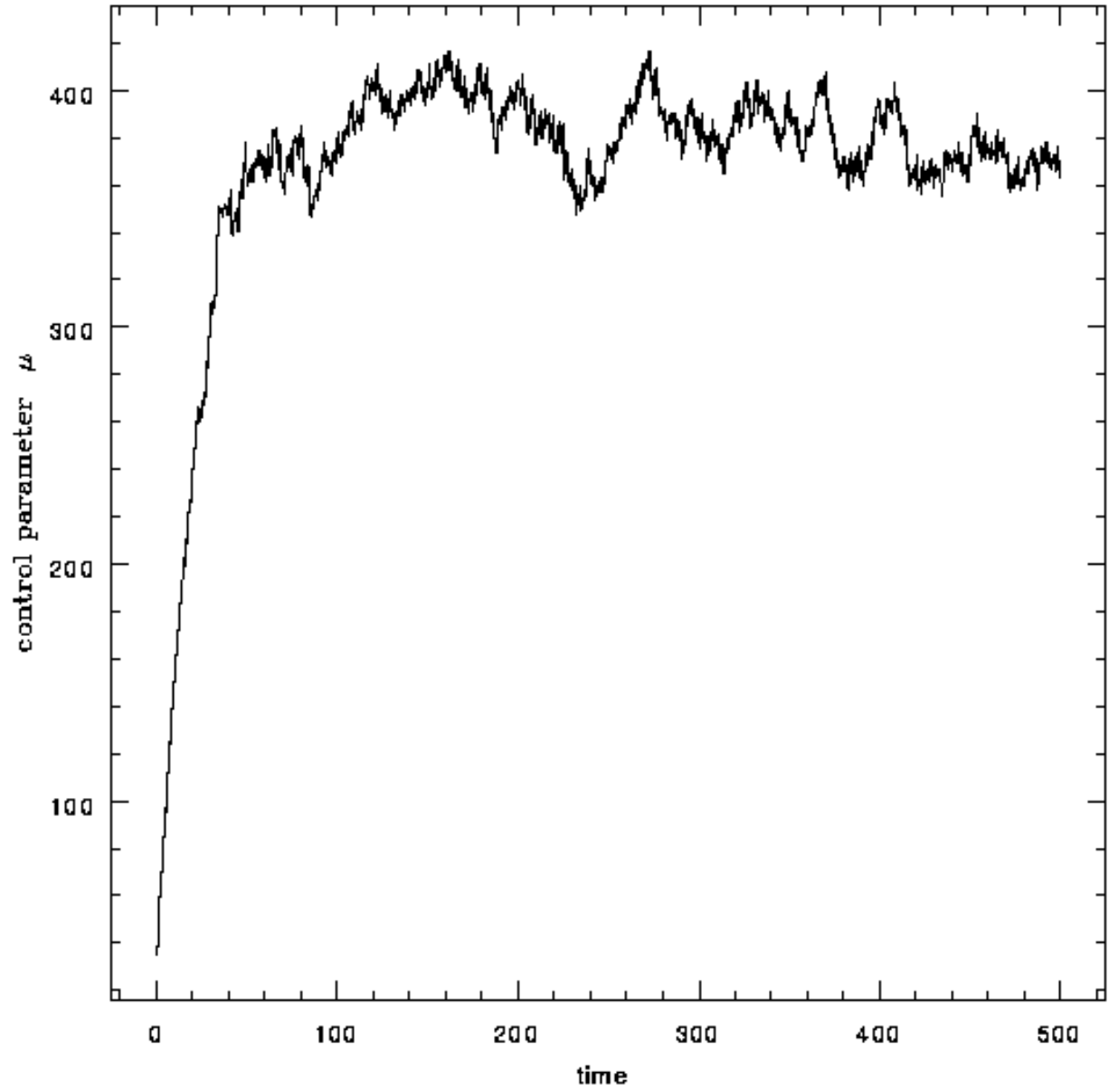
\end{document}